\documentclass[pre,showpacs]{revtex4-1}
\usepackage{bm}
\usepackage{graphicx}
\begin{document}

\title{Mass Dependence of Instabilities of an Oscillator with Multiplicative and Additive Noise}
\author{Moshe Gitterman}
\affiliation{Department of Physics, Bar-Ilan University, Ramat-Gan, IL52900 Israel}
\author{David A. Kessler}
\affiliation{Department of Physics, Bar-Ilan University, Ramat-Gan, IL52900 Israel}

\begin{abstract}
We study the instabilities of a harmonic oscillator subject to  additive and dichotomous multiplicative noise, focussing on the dependance of the instability threshold on the mass. 
For multiplicative noise in the damping, the instability threshold is crossed as the mass is decreased, as long as the smaller damping is in fact negative.  For multiplicative noise in the stiffness, the situation is more complicated and in fact the transition is reentrant for intermediate noise strength and damping.  For multiplicative noise in the mass, the results depend on the implementation of the noise.  One can take the velocity or the momentum to be conserved as the mass is changed. In these cases increasing the mass destabilizes the system. Alternatively, if the change in mass is caused by the accretion/loss of particles to the Brownian particle, these processes are asymmetric with momentum conserved upon accretion and velocity upon loss.  In this case, there is no instability, as opposed to the other two implementations. We also study the distribution of the energy, finding a power-law cutoff at a value which increases with time.
\end{abstract}
\maketitle

\section{Introduction}

The harmonic oscillator is, of course, one of the most basic models in physics, and the noisy harmonic oscillator is a touchstone of statistical physics.  The simplest model of a noisy oscillator is that of an oscillator driven by additive, white noise:
\begin{equation}
m\frac{d^{2}x}{dt^{2}}+2\gamma \frac{dx}{dt}+ k x=\eta \left( t\right)
\label{1}
\end{equation}%
with white noise $\eta \left( t\right) $,%
\begin{equation}
\langle \eta(t) \rangle = 0, \qquad\qquad \langle \eta \left( t_{1}\right) \eta \left( t_{2}\right) \rangle=S \delta \left(
t_{2}-t_{1}\right)  \label{2}
\end{equation}
which, as is well known, leads to a stationary Gaussian distribution for $x$ and $v=dx/dt$, whose width is proportional to $\sqrt{S}$.  It is also possible to add noise, here multiplicative noise, via external modification of one of the system parameters.  One such a model is that  of  a harmonic oscillator with random damping strength, $\gamma(t)$:
\begin{equation}
m\frac{d^2x}{dt^2} + 2\gamma(1+\xi(t))\frac{dx}{dt} + kx = 0
\end{equation}
with $\xi(t)$ another white noise.  This equation was first used~\cite{west} to analyze water waves influenced by a turbulent wind field.  It also can be used to describe,
upon replacing $x$ and $t$ by the order parameter and spatial coordinate, respectively, phase transition dynamics in a moving system~\cite{git1}.  Here the first derivative term represents advection by some external flow.  In this guise, the equation has been used to study phase transitions under shear~\cite{25}, open flows of liquids~\cite{26}, dendritic growth~\cite{28}, chemical waves~\cite{29} and motion of vortices~\cite{30}.

Another possibility is to randomly modulate the stiffness parameter, $k$, so that
\begin{equation}
m\frac{d^2x}{dt^2} + 2\gamma\frac{dx}{dt} + k(1+\xi(t))x = 0 .
\end{equation}
There are also many applications of this model in different fields in physics,
such as wave propagation in a random medium \cite{ishi}, spin precession in
a random external field \cite{kubo}, turbulent flow on the ocean surface 
\cite{phil}, and as well as in biology (population dynamics \cite{tur}), in
economics (stock market prices \cite{tak}) and so on.

Recently, one of us~\cite{gi,gi1,gi2,gi3} considered yet another way of introducing multiplicative noise, namely via a fluctuating mass term.  Here, as opposed to the other two possibilities outlined above, the mass term must always be positive, so white noise is ruled out.  One possibility is dichotomous noise, where the mass alternates randomly between two values, $m_\pm$. so that the equation reads
\begin{equation}
m_\pm\frac{d^2x}{dt^2} + 2\gamma\frac{dx}{dt} + k(1+\xi(t))x = 0 .
\end{equation}
This model was originally conceived off in the context of a Brownian particle undergoing random adsorption and desorption.  Many other applications of an oscillator
with a random mass have been considered~\cite{lam}, including ion-ion reactions \cite{gad}-\cite%
{gad1}, electrodeposition \cite{per}, granular flow \cite{gol}-\cite{see},
cosmology \cite{benz}-\cite{weid}, film deposition \cite{kai}, traffic jams 
\cite{nag}-\cite{benn}, and the stock market \cite{aus}-\cite{aus1}.

The most striking phenomenon of these models with multiplicative noise is the existence of transitions, with no steady state for sufficiently large noise.  It has been traditional in these studies of harmonic oscillators subject to multiplicative noise to scale out the mass, setting it to unity.  This is of course possible, and one can write the critical noise amplitude (which is dimensionless) in terms of the dimensionless parameters $k/(m\lambda^2)$ and $\gamma/(m\lambda)$, where $\lambda$ is the inverse correlation time of the noise. (We will discuss the parallel situation in the white noise limit below).  However, doing so obscures an important aspect of the physics of the problem; namely, the nontrivial dependence on the mass. Thus, the mass is an important parameter in a way that it is not in the presence of additive noise only.  One major goal of this work is  to consider the mass dependence explicitly. We will see that in some cases the mass dependence is even nonmonotonic.   In addition, we explore some other features of the problem which seem to have escaped notice.  Primary among them is the fact that the statistics of the observables, $x$, $v$ and $E$ are anomalous, with power-law tails. As opposed to models with restabilization at  a nonlinear fixed point~\cite{yamada}, here there is no cutoff of the power-law at long times.  This is related also to the scale invariance of the harmonic oscillator.  Thus, even for parameters for which $\langle E\rangle$, say, is finite, higher moments of $E$ are infinite in the infinite-time limit. They are finite at finite time due to the presence of a time-dependent cutoff on the power-law distribution, so that the moments grew exponentially in time.  This behavior is also reflected in a single time trace, with the dynamics having a bursty, intermittent character~\cite{yamada}.  The exponent governing the tail of the distribution determines which moments have finite infinite-time limits. Thus the power-law nature of the distribution is intimately related to the infinite series of  transitions in the model,~\cite{Arnold} corresponding to the onset of convergence of higher and higher moments. 

The plan of the paper is as follows.  We first discuss the mass dependence of the well-known transition corresponding to the divergence of the average energy (equivalently $\langle x^2 \rangle$). We first discuss the case of random damping, where most of the general features are already present. We then move on to treat the cases of random stiffness and random mass in turn.  In the second section, we discuss the distribution of the energy and their connection to the infinite set of transitions for different moments. In the last section we present our conclusions.

\section{The Mass Dependence of the Energy Instability}
\subsection{Random Damping}
We first discuss the case of random damping. We work in the framework of symmetric dichotomous noise, recovering the white-noise case by taking the small correlation time, large amplitude limit.  Thus the damping $\gamma$ switches randomly in a Poisson fashion between two values, $\gamma_\pm$.  If both $\gamma_\pm$ are positive, there is of course no instability.  However, a negative $\gamma_-$ by itself is not enough to ensure instability. The basic starting point of the analysis is the equations of motion of the second moments.  These moment equation can be derived either from the Langevin equation or from a Fokker-Planck formalism.  The former method was presented, for example, in Ref. \cite{gi,gi1,gi2,gi3}.  In the Fokker-Planck formalism, we construct  two coupled Fokker-Planck equations for the probability density $P_\pm(x,v)$, conditioned on the value of $\gamma$.  The equations read
\begin{eqnarray}
\dot{P}_+ &=& \frac{D}{m^2}\frac{\partial^2}{\partial v^2} P_+ + \frac{2\gamma_+}{m} \frac{\partial}{\partial v} (vP_+) + \frac{kx}{m}\frac{\partial}{\partial v} P_+ - v\frac{\partial}{\partial x} P_+ + \frac{\lambda}{2}\left( P_+ - P_-\right) \nonumber\\
\dot{P}_- &=& \frac{D}{m^2}\frac{\partial^2}{\partial v^2} P_- + \frac{2\gamma_-}{m} \frac{\partial}{\partial v} (vP_-) + \frac{kx}{m}\frac{\partial}{\partial v} P_- - v\frac{\partial}{\partial x} P_- - \frac{\lambda}{2}\left( P_+ - P_-\right)
\end{eqnarray}
Here, the diffusion constant $D$ is related to the strength of the additive noise and $\lambda$ is the decay rate of the correlator of the dichotomous noise, and is inversely proportional to the average switching time, $\tau$, by $\lambda=2/\tau$.  From these two equations, it is easy to generate a closed set of moment equations for the six conditioned quantities, $\langle x^2 \rangle_\pm$, $\langle v^2 \rangle_\pm$ and $\langle xv \rangle_\pm$.  The equations read
\begin{eqnarray}
\frac{d}{dt}\langle x^2 \rangle_\pm &=& 2 \langle xv \rangle_\pm \mp \frac{\lambda}{2} \left (\langle x^2 \rangle_+ - \langle x^2 \rangle_-\right) \nonumber \\
\frac{d}{dt}\langle xv \rangle_\pm &=&  - \frac{k}{m} \langle x^2 \rangle_\pm - \frac{2}{m}\gamma_\pm \langle xv \rangle_\pm +  \langle v^2 \rangle_\pm \mp \frac{\lambda}{2} \left (\langle xv \rangle_+ - \langle xv \rangle_- \right)\nonumber \\
\frac{d}{dt}\langle v^2 \rangle_\pm &=& \frac{2D}{m^2} -  \frac{2k}{m} \langle xv \rangle_\pm - \frac{4}{m}\gamma_\pm \langle v^2 \rangle_\pm \mp \frac{\lambda}{2} \left (\langle v^2 \rangle_+ - \langle v^2 \rangle_- \right)
\label{sys}
\end{eqnarray}
These are, of course, equivalent to the  moment equations arising from the Langevin equation.
Setting the time derivatives to zero and solving yields
\begin{equation}
\langle x^2 \rangle = \frac{1}{2}\left(\langle x^2 \rangle_+ + \langle x^2 \rangle_-\right) = \frac{D}{k} \frac{(\gamma_+ + \gamma_-)(4k + 3m\lambda^2) + 4\lambda (\gamma_+^2+\gamma_-^2) + m^2\lambda^3}{(\gamma_++\gamma_-)(m^2\lambda^3 + 4m\lambda k + 8\lambda\gamma_+\gamma_-) + (\gamma_+^2 + \gamma_-^2)m\lambda^2   + \gamma_+\gamma_-(10m\lambda^2 + 16k) }
\end{equation}
Setting the mass to unity yields the result presented in earlier works.  Alternatively, our present result can be reconstructed from the  standard, $m=1$ result by the substitutions 
\begin{equation}
\gamma_\pm \Rightarrow \frac{\gamma_\pm}{m} \ ; \qquad k \Rightarrow \frac{k}{m}  \ ; \qquad  D \Rightarrow \frac{D}{m^2}
\label{subs}
\end{equation}

It is clear that as long as $\gamma_\pm$ are both positive, $\langle x^2 \rangle$ is positive, and there is no instability.  It is convenient to rewrite $\langle x^2 \rangle$ in terms of $\gamma$ and $\sigma$, where $\gamma_\pm=\gamma (1 \pm \sigma)$, yielding
\begin{equation}
\langle x^2 \rangle = \frac{D}{2k\gamma}\frac{8\gamma k + 4km\lambda + 6m\gamma \lambda^2 + m^2\lambda^3 + 8\gamma^2\lambda(1+\sigma^2) }{
8\gamma k + 4km\lambda+6m\gamma\lambda^2+m^2\lambda^3+8\gamma^2\lambda - 4\gamma\sigma^2(m\lambda^2+2k+2\gamma\lambda)}
\end{equation}

Let us first consider the white noise limit:
\begin{equation}
\lambda \to \infty \ ; \qquad \sigma^2 \to \infty \ ; \qquad \frac{2\sigma^2}{\lambda} = S \ \textit{fixed}
\end{equation}
where $\langle x^2 \rangle$ reduces to 
\begin{equation}
\langle x^2 \rangle_\textit{\small{WN}} = \frac{D}{k(1-2S\gamma/m)}
\end{equation}
Here the critical noise amplitude is $S_c = m/(2\gamma)$, beyond which no steady-state exists.  This critical noise amplitude is proportional to the  mass, so that low mass systems are more unstable.

\begin{figure}
\includegraphics[width=0.7\textwidth]{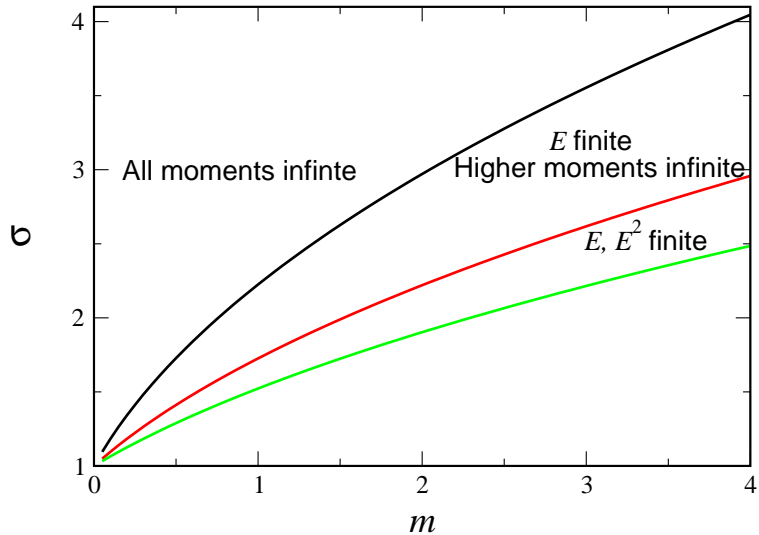}
\caption{Instability boundaries for the harmonic oscillator with random damping.  Above the uppermost line, all moments of the energy diverge. Between this line and the curve underneath, the first moment of the energy is finite, and all higher moments diverge.  Similarly, below the bottom line, the second moment of the energy is also finite. Below the bottom curve, the third moment of the energy is finite as well. Note that for $\sigma<1$, i.e. both $\gamma_\pm>0$, there is no instability. Here, $\gamma=0.025$, $k=1$, and $\lambda=0.2$.}
\label{gam_bound}
\end{figure}

Turning now to the general dichotomous system, we see that the numerator is positive definite, while the denominator is negative for sufficiently large $\sigma^2$.  The critical $\sigma^2$ is
\begin{equation}
\sigma^2_c = \frac{8\gamma k + 4km\lambda+6m\gamma\lambda^2+m^2\lambda^3+8\gamma^2\lambda}{4\gamma(m\lambda^2+2k+2\gamma\lambda)}
\end{equation}
It can be checked that $\sigma^2_c(m=0)=1$ and increases monotonically with $m$, diverging for large $m$ as $m\lambda/(4\gamma)$, in accord with our white noise limit result.  This implies that for any given $\sigma^2>1$, the system is unstable for small $m$ and is stable for $m>m_c(\sigma)$, where the critical $m$, $m_c$, increases with $\sigma$.  The instability boundary is indicated by the upper curve in Fig. \ref{gam_bound} (the other curves will be discussed below in Sec. II).  We thus see that for the dichotomous noise limit, the mass dependence is quite complicated.

As already pointed out by Kitahara, et al.~\cite{Kitahara}, an equivalent way to demonstrate the instability is to examine the dynamics of the system without external additive driving $\eta$.  Then, the system Eqs. (\ref{sys}) is a homogenous linear system characterized by a $6\times 6$ matrix.  As usual, the presence of positive eigenvalues corresponds to existence of an instability, which yields precisely the same stability criterion as given above.

\subsection{Random Spring Stiffness}
We now turn to the case where the spring stiffness, $k$, varies between two values, $k_\pm$.  The Fokker-Planck equation and the resulting moment equations are directly parallel to the previous case, and the steady-state solution is
\begin{equation}
\langle x^2 \rangle = 4D\frac{(m\lambda+2\gamma)\left[2(k_++k_-)+4\gamma\lambda+ m\lambda^2\right]}{(32\gamma m \lambda + 64\gamma^2) k_+k_-+(24\gamma^2m\lambda^2+4\gamma m^2\lambda^3+32\gamma^3\lambda)(k_++k_-)-(k_+-k_-)^2m^2\lambda^2}
\end{equation}
As before, this result can be recovered from the standard $m=1$ version by the substitutions given in Eq. (\ref{subs}).  The first thing to note is that, as opposed to the previous case, here the denominator can be negative even if $k_\pm$ are both positive.  This is due to the fact that changing for $k_-$ to $k_+$ increases the energy of the system, and even though the reverse transition decreases the energy, the two do not have to balance.  Rewriting things in terms of $k$ and $\sigma$, with $k_\pm\equiv k(1\pm \sigma)$, yields
\begin{equation}
\langle x^2 \rangle = \frac{D}{k}\frac{(m\lambda+2\gamma)\left(4k+4\gamma\lambda+ m\lambda^2\right)}{2\gamma(m\lambda + 2\gamma)(m\lambda^2+4\lambda\gamma+4k)-\sigma^2 k (m\lambda  + 4\gamma)^2}
\end{equation}
In the white noise limit, this reduces to
\begin{equation}
\langle x^2 \rangle_\textit{\small{WN}} = \frac{D}{ 2\gamma k (1 - kS/(4\gamma))}
\end{equation}
Thus in the white noise limit, the system is unstable for $S>4\gamma/k$ {\em independent} of the mass, as opposed to the case of the white noise random damping.  

\begin{figure}
\includegraphics[width=0.55\textwidth]{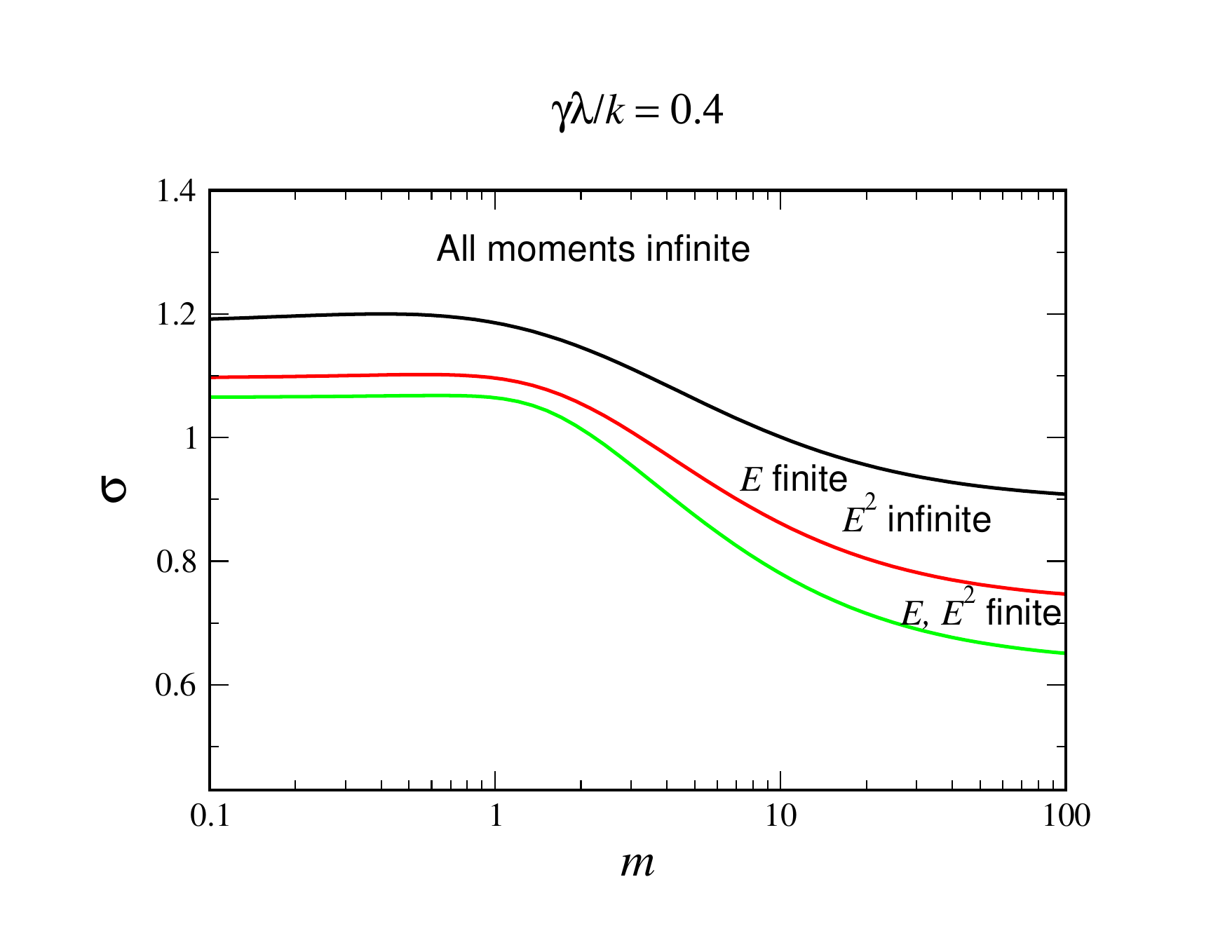}\hspace{-1.8cm}\includegraphics[width=0.55\textwidth]{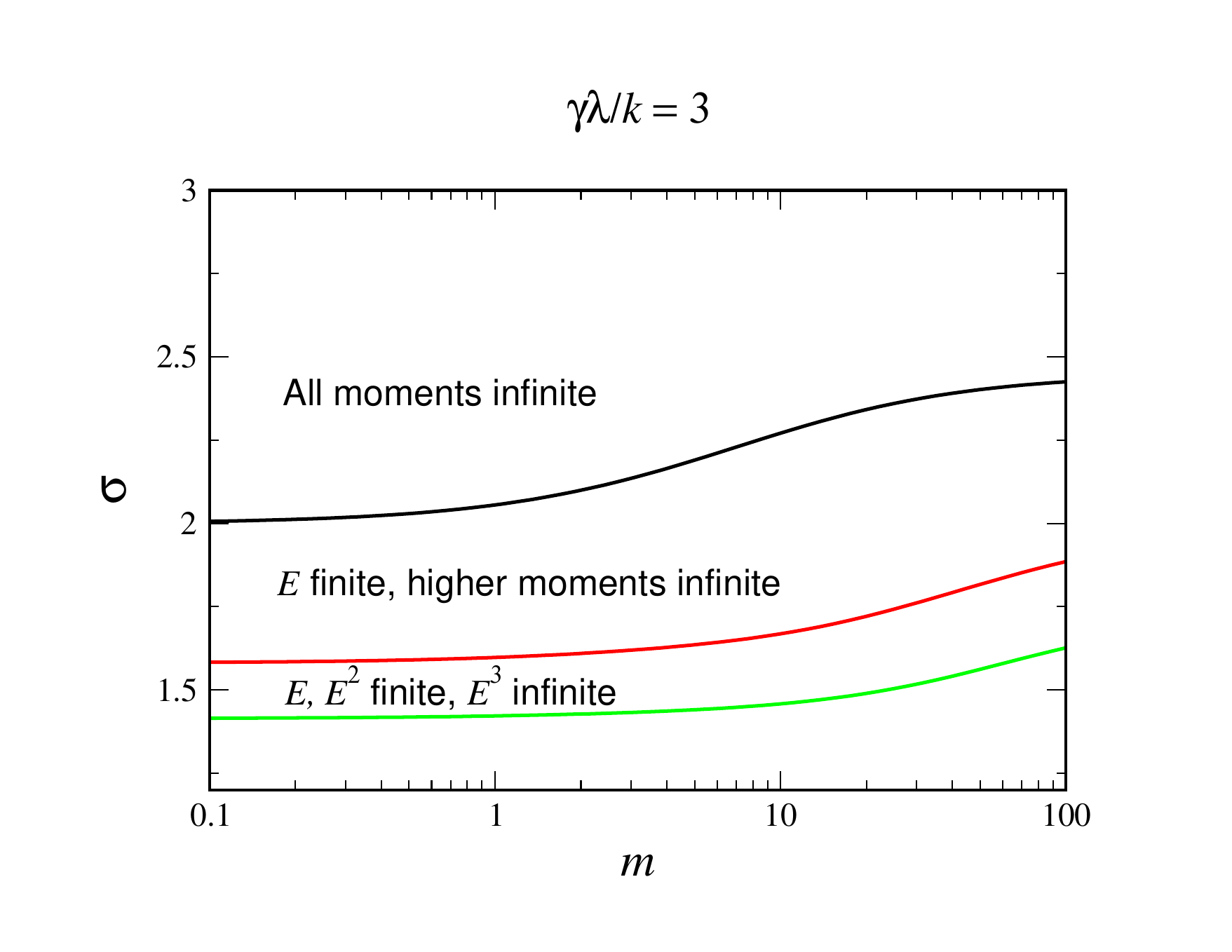}
\caption{Instability boundaries for the harmonic oscillator with random stiffness.  Above the uppermost line, all moments of the energy diverge. Between this line and the curve underneath, the first moment of the energy is finite, and all higher moments diverge.  Similarly, below the bottom line, the second moment of the energy is also finite. Below the bottom curve, the third moment of the energy is finite as well. Note that for in the left panel, with $\gamma$ small, there is an instability at large $m$ even for $\sigma<1$, i.e. both $k_\pm>0$. In the left panel, $\gamma=0.4$.  In the right panel, $\gamma=3$.  In both cases, $k=1$, and $\lambda=1$.}
\label{k_stable}
\end{figure}

Reverting to the general dichotomous case, we find that 
the critical $\sigma_c^2(m)$ above which the instability sets in is
\begin{equation}
\sigma_c^2 = \frac{2\gamma(m\lambda + 2\gamma)(m\lambda^2+4\lambda\gamma+4k)}{k (m\lambda  + 4\gamma)^2}
\end{equation}
In general, this function is not monotonic in $m$.  It has an extremum at 
\begin{equation}
m_\textit{\small{ext}} = \frac{4\gamma^2}{2k-\gamma\lambda}
\end{equation}
so that for $\gamma < 2k/\lambda$, it has a maximum at positive $m$, with the value
\begin{equation}
\sigma^2_\textit{\small{ext}} = \left[ 1 + \frac{\gamma\lambda}{2k} \right]^2
\end{equation}
In addition, note that 
\begin{equation}
\sigma^2_c(0) =  1 + \frac{\lambda\gamma}{k} ; \qquad\qquad \sigma^2_c(\infty) = \frac{2\gamma\lambda}{k}
\end{equation}
Thus, for $\gamma > 2k/\lambda$, $\sigma^2_c$ is monotonically increasing in $m$, whereas for $k/\lambda < \gamma < 2k/\lambda$, $\sigma_c^2$ rises with $m$ and then fails to a value greater than its value at 0, and for $\gamma < k/\lambda$, $\sigma^2_c$ rises with $m$ and then falls below its $m=0$ value.  The upshot of this is that for large damping, $\gamma > 2k/\lambda$, the system is stable for $\sigma^2 < \sigma^2_c(0)$, independent of mass, while for larger $\sigma^2$, the system is stable only above some critical value of $m$ which increases with $\sigma^2$.  For smaller damping, the situation is more complex.  For intermediate damping, $k/\lambda < \gamma < 2k/\lambda$, the system is always stable independent of $m$ for $\sigma^2 < \sigma_c^2(0)$.  For $\sigma_c^2(0) < \sigma^2 < \sigma_c^2(\infty)$, the system is stable above a critical mass which increases with $\sigma^2$.  For $\sigma^2_c(\infty) < \sigma^2 < \sigma^2_\textit{\small{ext}}$, the system is stable in a finite range of $m$, which shrinks from top and bottom as $\sigma^2$ increases. Finally, for $\sigma^2 > \sigma^2_\textit{\small{ext}}$, the system is always unstable, independent of $m$.  For $\gamma<k/\lambda$, on the other hand, the system is stable for $\sigma^2<\sigma^2(\infty)$, while in the very narrow band $\sigma^2(\infty)<\sigma^2<\sigma^2_\textit{\small{ext}}$, the system is stable for an intermediate range of $m$'s.  For $\sigma^2>\sigma^2_\textit{\small{ext}}$, the system is always unstable.  The cases of very small and large damping are shown in Figs. \ref{k_stable}.  It should also be noted that the system is unstable with $\sigma^2<1$, i.e. with both $k_\pm >0$, only for very weak damping,
$\gamma<k/(2\lambda)$, and for sufficiently large $m$, 
\begin{equation}
m>2\left(\frac{\gamma}{\lambda}\right)\frac{3\gamma\lambda + \sqrt{\gamma\lambda(4k+\gamma\lambda)}}{k-2\gamma\lambda}
\end{equation}

\subsection{Random Mass}

\begin{figure}
\includegraphics[width=0.55\textwidth]{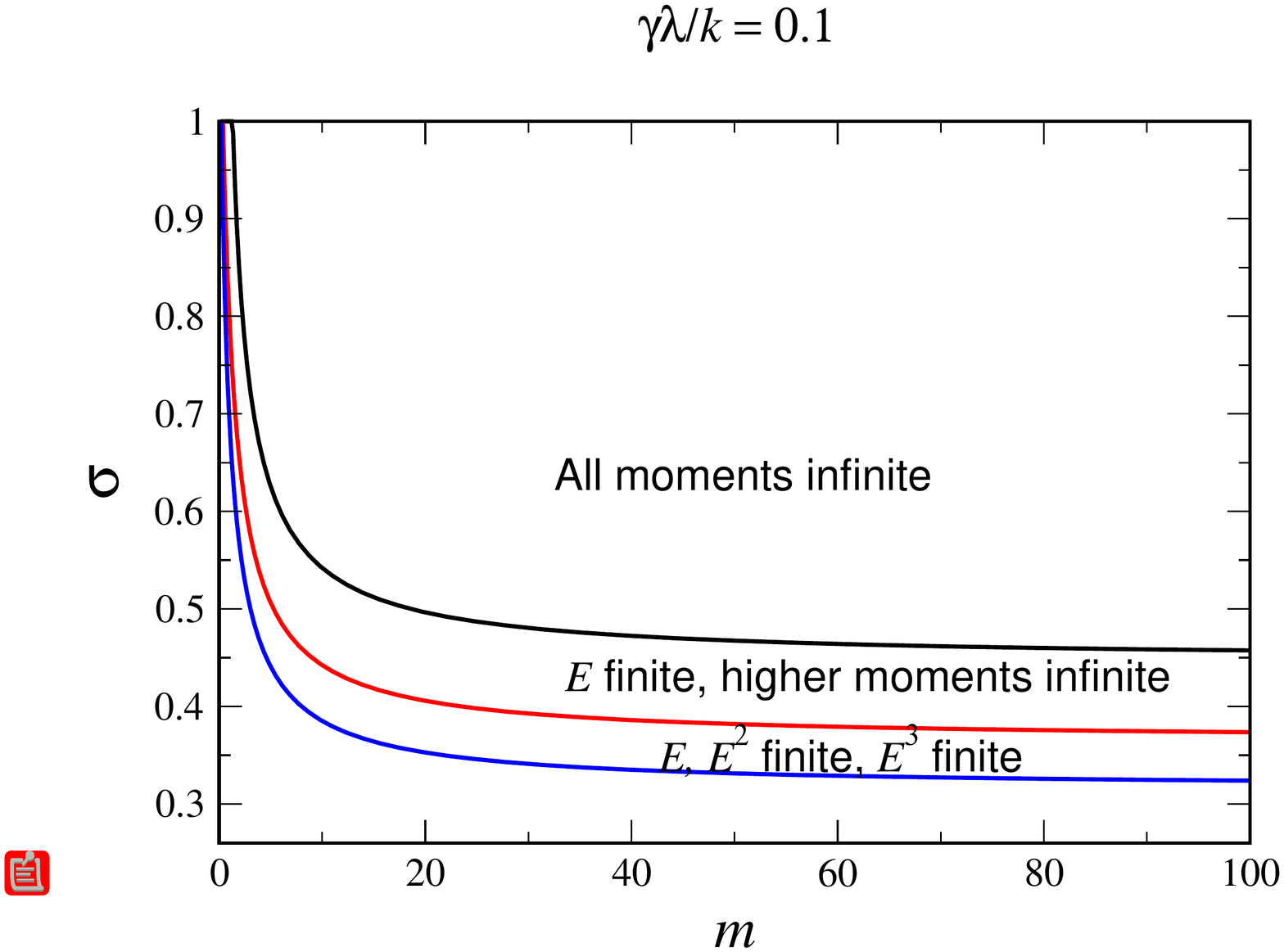}\hspace{-1.88cm}\includegraphics[width=0.55\textwidth]{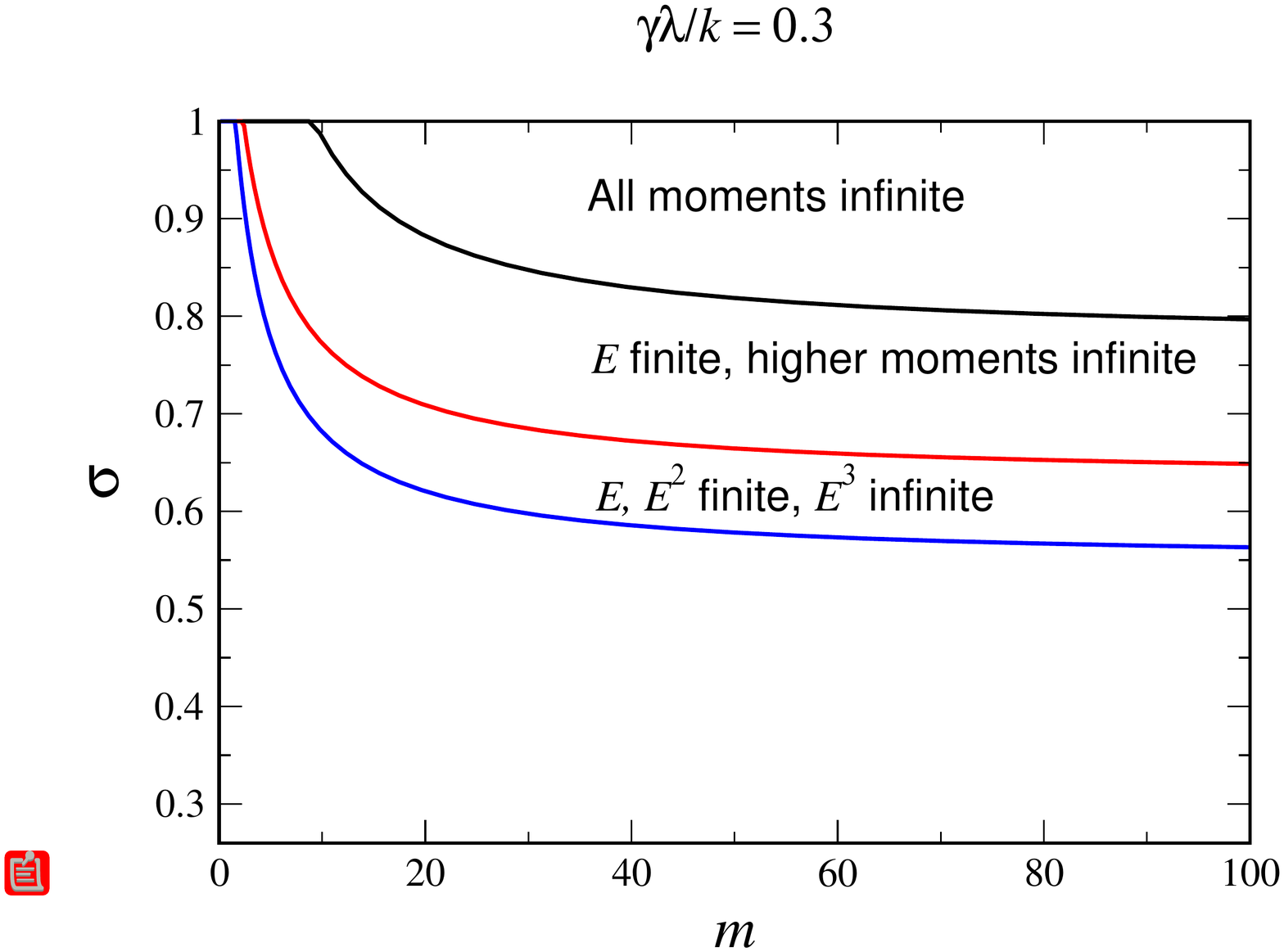}
\caption{Instability boundaries for the harmonic oscillator with random mass coefficient, and velocity (or momentum, the phase boundaries coincide in the two cases).  Above the uppermost line, all moments of the energy diverge. Between this line and the curve underneath, the first moment of the energy is finite, and all higher moments diverge.  Similarly, below the bottom line, the second moment of the energy is also finite. Below the bottom curve, the third moment of the energy is finite as well. Here the model is physical only for $\sigma<1$, i.e. both $m_\pm>0$, there is no instability. In the left panel, $\gamma=0.1$.  In the right panel, $\gamma=0.3$ Here,  $k=1$, and $\lambda=1$.}
\label{m_sigc}
\end{figure}

Recently~\cite{gi,gi1,gi2,gi3}, a model in which the mass term randomly fluctuates was studied.  In this case, it is clearly important that the fluctuations do not change the sign of the mass term, which rules out the possibility of white noise.  Nevertheless, it is possible to consider the case of dichotomous noise where the ``mass" fluctuates between two values, $m_\pm$.  Depending on the system being modeled, the random ``mass" can represent, e.g., the inductance in an LRC circuit, the moment of inertia in a physical pendulum, etc.  Depending on the physical application, different mathematical models are called for.  For example, in the case of a random inductance, the charge and current are conserved at the moment of the fluctuation.  For the case of a random moment of inertia, one would expected the angle and angular momentum to be conserved at the moment of the fluctuation.  Thus, these two cases lead to two different mathematical models, one which reads
\begin{equation}
m_\pm \frac{d^2}{dt^2} x = -\gamma \dot{x} - k x + \eta
\end{equation}
and the second which reads
\begin{equation}
\frac{d}{dt}(m_\pm \dot x) = -\gamma \dot{x} - kx + \eta
\end{equation}
In the first, continuous velocity, case, which was the case studied in Refs. \cite{gi,gi1,gi2,gi3}, the moment equations read
\begin{eqnarray}2
\frac{d}{dt}\langle x^2 \rangle_\pm &=& 2 \langle xv \rangle_\pm \mp \frac{\lambda}{2} \left (\langle x^2 \rangle_+ - \langle x^2 \rangle_-\right) \nonumber \\
\frac{d}{dt}\langle xv \rangle_\pm &=&  - \frac{k}{m_\pm} \langle x^2 \rangle_\pm - \frac{2\gamma}{m_\pm} \langle xv \rangle_\pm +  \langle v^2 \rangle_\pm \mp \frac{\lambda}{2} \left (\langle xv \rangle_+ - \langle xv \rangle_- \right)\nonumber \\
\frac{d}{dt}\langle v^2 \rangle_\pm &=& \frac{2D}{m_\pm^2} -  \frac{2k}{m_\pm} \langle xv \rangle_\pm - \frac{4\gamma}{m_\pm} \langle v^2 \rangle_\pm \mp \frac{\lambda}{2} \left (\langle v^2 \rangle_+ - \langle v^2 \rangle_- \right)
\label{sys1}
\end{eqnarray}
The solution is
\begin{equation}
\langle x^2\rangle = \frac{2D}{km_+m_-}\frac{\lambda(k + \gamma\lambda)(m_+ + m_-)^3 + 16\gamma(k+\gamma\lambda) m_+m_- + \gamma\lambda^2 2 m_+m_-(m_+ + m_-) + \lambda^3 m_+m_-(m_+^2+m_-^2)}{16\gamma k \lambda(m_++m_-)+64\gamma^2(k+\gamma\lambda) + 24\gamma^2\lambda^2(m_++m_-)-k\lambda^2(m_+^2+m_-^2) + 2\gamma\lambda^3(m_+ + m_-)^2}
\label{randmass1}
\end{equation}
As opposed to the cases of random damping and random stiffness considered above, here for physical reasons the mass must always be positive, and so we cannot pass to the white noise limit. Examining the dichotomous result, Eq. (\ref{randmass1}),
the numerator is positive definite for $m_\pm >0$, while the denominator vanishes at a critical noise strength, $\sigma$, where $m_\pm = m(1\pm\sigma)$:
\begin{equation}
\sigma_c^2 = \frac{2\gamma(2\gamma+m\lambda)(4\gamma\lambda+4k+ m\lambda^2)}{m^2 k \lambda^2}
\end{equation}
It is clear that $\sigma_c^2$ is a monotonically decreasing function of $m$, approaching $2\gamma\lambda/k$ as $m$ gets large.  Since the positivity of $m_\pm$ requires that $\sigma^2<1$, there is clearly no instability if $2\gamma\lambda>k$. If $2\gamma\lambda < k$, then there is an instability above some critical value of $m$, if $\sigma^2$ is large enough such that $\sigma_c^2(m)<\sigma^2<1$.  Turning this around, for fixed $1>\sigma^2>2\gamma\lambda/k$, there is an instability for $m$ greater than a critical value.  For $m<\gamma/\lambda(\sqrt{1+16r+32r^2}-3-4r)/(1-r)$, where $r=k/(2\gamma\lambda)<1$, there is no instability for any $\sigma^2<1$.  This is illustrated in Fig. \ref{m_sigc}.

For the second case, with momentum conserved at the transition, the moment equations read
\begin{eqnarray}
\frac{d}{dt}\langle x^2 \rangle_\pm &=& 2 \langle xv \rangle_\pm \mp \frac{\lambda}{2} \left (\langle x^2 \rangle_+ - \langle x^2 \rangle_-\right) \nonumber \\
\frac{d}{dt}\langle xv \rangle_\pm &=&  - \frac{k}{m_\pm} \langle x^2 \rangle_\pm - \frac{2\gamma}{m_\pm} \langle xv \rangle_\pm +  \langle v^2 \rangle_\pm - \frac{\lambda}{2} \left (\langle xv \rangle_\pm - \frac{m_\mp}{m_\pm}\langle xv \rangle_\mp \right)\nonumber \\
\frac{d}{dt}\langle v^2 \rangle_\pm &=& \frac{2D}{m_\pm^2} -  \frac{2k}{m_\pm} \langle xv \rangle_\pm - \frac{4\gamma}{m_\pm} \langle v^2 \rangle_\pm - \frac{\lambda}{2} \left (\langle v^2 \rangle_\pm - \left(\frac{m_\mp}{m_\pm}\right)^2\langle v^2 \rangle_\mp \right)
\label{sys2}
\end{eqnarray}
One way to see that these are the correct equations is to write the moment equations for the variables \{$\langle x^2 \rangle_\pm,\ \langle xp \rangle_\pm,\ \langle p^2 \rangle_\pm$\} where $p_\pm = m_\pm v_\pm$.  In this case, the solution for $\langle x^2 \rangle$ is
\begin{equation}
\langle x^2 \rangle = \frac{D}{k}\frac{(4\gamma+\lambda(m_+ + m_-))(8k + 8\gamma\lambda + \lambda^2(m_+ + m_-))}{64\gamma^2(k+\gamma\lambda) + (m_++m_-)(16k\gamma \lambda +24\gamma^2\lambda^2) + (m_+^2 + m_-^2)(2\gamma\lambda^3 - k\lambda^2) + m_+m_-(4\gamma\lambda^3 +2\lambda^2)}
\end{equation}
As expected, the numerator is positive definite for $m_\pm>0$, while the denominator is negative if
\begin{equation}
\sigma^2 > \frac{2\gamma(2\gamma+m\lambda)(4\lambda\gamma + 4k+ m\lambda^2)}{m^2k\lambda^2}
\end{equation}
This is exactly the same condition for instability as for the conserved velocity model, even though $\langle x^2 \rangle$ is different in the two models.  The same is true for the instability conditions for the higher-order moments.

There is yet a third possible model.  If the changes in the ``mass" term represent actual changes in the mass due to accretion/desorption of particles to the Brownian particle, then due to Newton's third law, the addition of mass conserves momentum, whereas the loss of mass conserves velocity.  Then the moment equations read
\begin{eqnarray}
\frac{d}{dt}\langle x^2 \rangle_\pm &=& 2 \langle xv \rangle_\pm \mp \frac{\lambda}{2} \left (\langle x^2 \rangle_+ - \langle x^2 \rangle_-\right) \nonumber \\
\frac{d}{dt}\langle xv \rangle_+ &=&  - \frac{k}{m_+} \langle x^2 \rangle_+ - \frac{2\gamma}{m_+} \langle xv \rangle_+ +  \langle v^2 \rangle_+ - \frac{\lambda}{2} \left (\langle xv \rangle_+ - \frac{m_-}{m_+}\langle xv \rangle_- \right)\nonumber \\
\frac{d}{dt}\langle xv \rangle_- &=&  - \frac{k}{m_-} \langle x^2 \rangle_- - \frac{2\gamma}{m_-} \langle xv \rangle_- +  \langle v^2 \rangle_- - \frac{\lambda}{2} \left (\langle xv \rangle_- - \langle xv \rangle_+ \right)\nonumber \\
\frac{d}{dt}\langle v^2 \rangle_+ &=& \frac{2D}{m_+^2} -  \frac{2k}{m_+} \langle xv \rangle_+ - \frac{4\gamma}{m_+} \langle v^2 \rangle_+ - \frac{\lambda}{2} \left (\langle v^2 \rangle_+ - \left(\frac{m_-}{m_+}\right)^2\langle v^2 \rangle_- \right) \nonumber\\
\frac{d}{dt}\langle v^2 \rangle_- &=& \frac{2D}{m_-^2} -  \frac{2k}{m_-} \langle xv \rangle_- - \frac{4\gamma}{m_-} \langle v^2 \rangle_- - \frac{\lambda}{2} \left (\langle v^2 \rangle_- - \langle v^2 \rangle_+ \right)
\label{sys3}
\end{eqnarray}
In this model, the Brownian particle losses kinetic energy both upon accretion and loss and thus the pumping mechanism present in the other versions is absent.  Indeed, solving the FPEs shows that the system does not possess an instability for any $m_\pm >0$!

\section{The Distribution of Energy and Higher Transitions}

\subsection{Random Damping}

\begin{figure}
\includegraphics[width=0.7\textwidth]{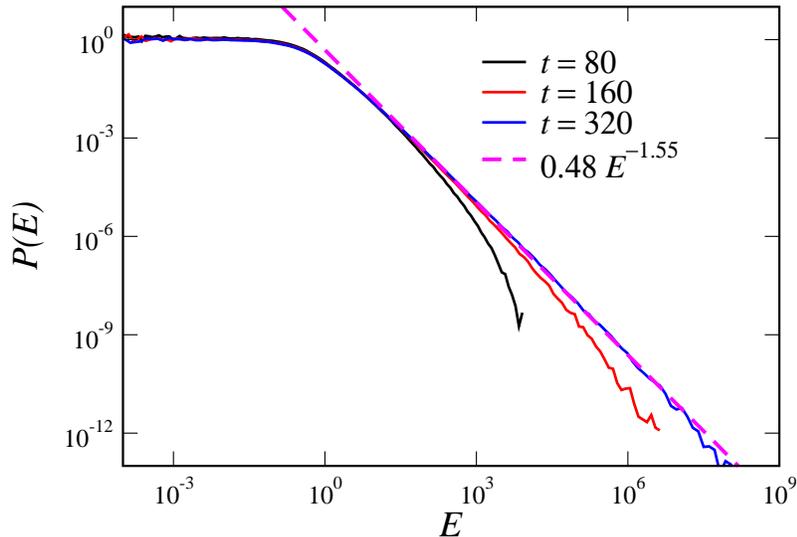}
\caption{The probability density of $E$ for the harmonic oscillator with random damping for various times, as measured by simulation.  Notice that as $t$ increases, the large $E$ behavior approaches a pure power-law, whereas for shorter times, it is cut-off.  Here $k=1$, $\lambda=0.2$, $m=1$, $\gamma=0.025$}
\label{power_m1}
\end{figure}

\begin{figure}
\includegraphics[width=0.7\textwidth]{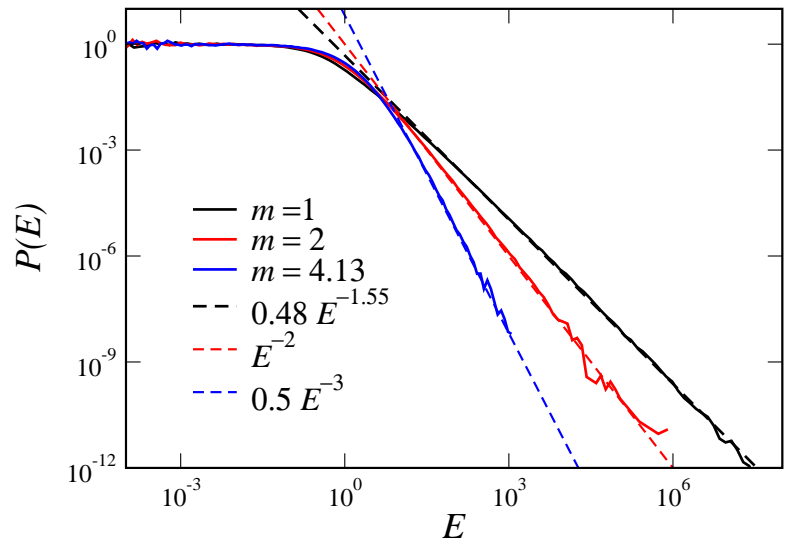}
\caption{The probability density of $E$ for the harmonic oscillator with random damping for various values of $m$, as measured by simulation.  Here, $k=1$, $\lambda=0.2$,  $\gamma=0.025$, and $t=640$.}
\label{gam_manym}
\end{figure}

Examining the how the distribution of $E=(kx^2+mv^2)/2$ changes with time is very instructive.  We see in Fig. \ref{power_m1} that the distribution is characterized by a power-law tail which is cut off beyond some value of $E$, a value which increases with time.  Given that in the absence of such a cutoff the mean value of $E$ would diverge (the power is smaller than 2 in magnitude), the mean is strongly time-dependent, as the moment equations predict. A similar phenomenon happens for a random walk in a logarithmic potential~\cite{KesBar}, where the distribution of position also has a time-dependent cutoff (that in that case grows as the square-root of time). As we lower the noise amplitude, or increase the mass, the power increases in magnitude.  The critical noise amplitude, or mass, is that value for which the power is equal to -2, at which point the mean value of $E$ is finite in the $t \to \infty$ limit, when the power-law is not cut off.  We see this in Fig. \ref{gam_manym}, where the power is close to -2 for $m=2$, as this $m$ is very close to the critical $m_c=2.045$ for these parameters.  This implies that just beyond the critical mass, the mean value of $E$ converges but the mean value of $E^2$ diverges.  This can be verified directly by looking at the moment equations for the fourth order moments, \{$\langle x^4\rangle_\pm,\ \langle x^3v\rangle_\pm,\ \langle x^2v^2\rangle_\pm,\ \langle xv^3\rangle_\pm,\ \langle v^4\rangle_\pm$\}. These moment equations read:
\begin{eqnarray}
\frac{d}{dt}\langle x^4 \rangle_\pm &=& 4 \langle xv^3 \rangle_\pm \mp \frac{\lambda}{2} \left (\langle x^4 \rangle_+ - \langle x^4 \rangle_-\right) \nonumber \\
\frac{d}{dt}\langle x^3v \rangle_\pm &=&  - \frac{k}{m} \langle x^4 \rangle_\pm - \frac{2}{m}\gamma_\pm \langle x^3v \rangle_\pm +  3\langle x^2v^2 \rangle_\pm \mp \frac{\lambda}{2} \left (\langle x^3v \rangle_+ - \langle x^3v \rangle_- \right)\nonumber \\
\frac{d}{dt}\langle x^2v^2 \rangle_\pm &=& \frac{2D}{m^2} - 2\frac{k}{m} \langle x^3 v \rangle_\pm - \frac{4}{m}\gamma_\pm \langle x^2v^2 \rangle_\pm +  2\langle xv^3 \rangle_\pm \mp \frac{\lambda}{2} \left (\langle x^2v^2 \rangle_+ - \langle x^2v^2 \rangle_- \right)\nonumber \\
\frac{d}{dt}\langle xv^3 \rangle_\pm &=& \frac{6D}{m^2}\langle xv\rangle_\pm - 3\frac{k}{m} \langle x^2 v^2 \rangle_\pm - \frac{6}{m}\gamma_\pm \langle xv^3 \rangle_\pm +  \langle v^4 \rangle_\pm \mp \frac{\lambda}{2} \left (\langle x^3v \rangle_+ - \langle x^3v \rangle_- \right)\nonumber \\
\frac{d}{dt}\langle v^4 \rangle_\pm &=& \frac{12D}{m^2}\langle x^2\rangle_\pm -  \frac{4k}{m} \langle xv^3 \rangle_\pm - \frac{8}{m}\gamma_\pm \langle v^4 \rangle_\pm \mp \frac{\lambda}{2} \left (\langle v^4 \rangle_+ - \langle v^4 \rangle_- \right)
\label{sys4}
\end{eqnarray}
These fourth-order moment equations are driven by the second-order moments.  To locate the transition, it is preferable to look at the eigenvalues of the $10\times 10$ stability operator.
The critical value of $m$ for which the fourth-order moments exists is $m_c^4=4.132$. Thus, we expect that the asymptotic slope of $P(E)$ for $m=4.13$ should be $-3$, as is also seen in Fig. \ref{gam_manym}. Thus, there is no steady-state solution of the fourth-order moment equations even for values of $m$ for which a steady-state solution of the second-order moment system exists.  Similarly, by looking at the sixth order moments, the condition for the existence of a steady-state solution is even more restrictive.  This is shown in Fig. \ref{gam_bound}, where the critical lines for the existence of stable solutions of the second, fourth and sixth moment equations are shown, corresponding to the borders of the regions of finiteness of $\langle E \rangle$, $\langle E^2 \rangle$ and $\langle E^3 \rangle$.  The critical $\sigma$ for the $n$th moment starts off linearly from $\sigma=1$ for $m=0$, turning over and behaving as
\begin{equation}
\sigma_c^n \approx \sqrt{\frac{2\lambda m}{3n+2}}
\end{equation}
for large $m$.

\begin{figure}
\includegraphics[width=0.7\textwidth]{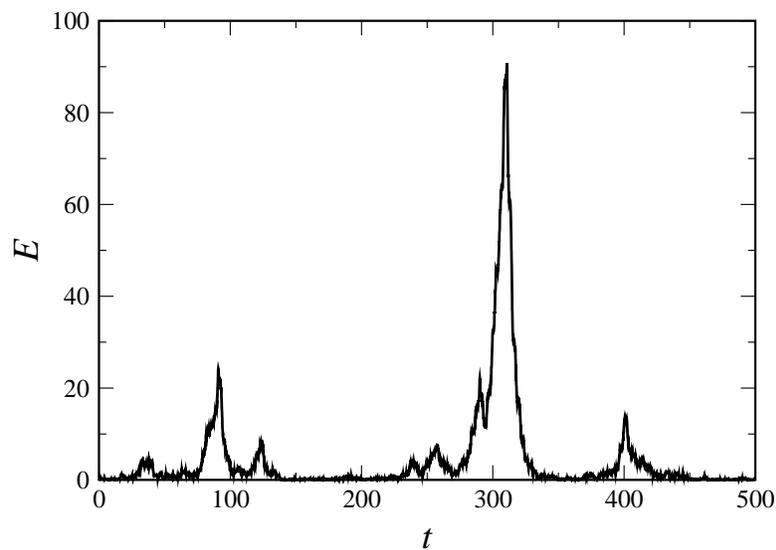}
\caption{Time trace for a single run of the harmonic oscillator with random damping, showing the intermittent behavior.  Here, $k=1$, $\lambda=0.2$, $m=1$, $\gamma=0.025$, $\sigma=3$.}
\label{g_burst}
\end{figure}

These anomalous statistics for $E$ imply that the system has an intermittent, bursty character.  We see this in Fig. \ref{g_burst}, where we plot $E(t)$ for a single long run.  The figure demonstrates clearly that the system is quiescent for a while, then undergoes after some random wait a large fluctuation, and returns to its quiescent state.

\subsection{Random Spring Stiffness }
As with the case of random damping, here also the fourth order moments are unstable for a wider range of parameters than the second order moment system.  Likewise, the sixth order moments are unstable for yet a wider range of parameters.  This is shown in Fig. \ref{k_stable}. The $m=0$ value of the critical $\sigma$ for the general $n$th order system is
\begin{equation}
(\sigma_c^n)^2(m=0) = 1 + \frac{2\gamma \lambda}{nk}
\end{equation}
and the $m\to\infty$ limit is
\begin{equation}
(\sigma_c^n)^2(m\to\infty) = \frac{8\gamma\lambda}{(n+2)k}
\end{equation}
 The $n$ dependence of the stability criterion is again indicative of the cut-off power-law distribution of the moments.

It should be noted that while the natural oscillation frequency of the spring, $\sqrt{k/m}$ varies between two values, the above model is not at all equivalent to Kubo's model~\cite{Kubo} of an oscillator with random frequency.  In fact, Kubo in his paper introduced two models, one of which is described by the Kubo equation
\begin{equation}
\dot{\phi} = \omega(t)
\end{equation}
where $\omega(t)=\omega_0 + \eta$, and $\eta$ is a fluctuating noise term.  Here, there is only a single degree of freedom, the angular one, and so clearly there can be no instability in the amplitude of oscillation.  Rather, there is simply a decay in time of the phase correlation.  Kubo also introduced a model with two degrees of freedom
\begin{eqnarray}
\dot{x} &=& - \omega(t) y \nonumber\\
\dot{y} &=& \omega(t) x
\end{eqnarray}
where $\omega$ is a noisy frequency centered on $\omega_0$.  Again here, there is no amplitude instability; rather the amplitude is exactly neutrally stable,
as can be seen by the fact that $x+iy$ can be exactly solved for as $x+iy=A e^{i\int \omega(t') dt'}$ for some complex constant $A$.  The lack of an instability can also be demonstrated for the dichotomous noise case by computing the eigenvalues of the Fokker-Planck operator for the moment system ${\langle x^2\rangle_\pm,\ \langle xy\rangle_\pm,\ \langle y^2\rangle_\pm} $, and verifying that they are all negative, except for an exact zero model corresponding to changes in the amplitude.  Thus, in this case, without damping, the system is exactly marginal, whereas the random stiffness model is always unstable in the absence of damping.

\subsection{Random Mass}

Again, the fourth-order system is unstable for a wider range of parameters, and the sixth-order system for a yet wider range. For example,
\begin{equation}
(\sigma_c^n)^2(m\to \infty) = \frac{8\gamma\lambda}{(n+2)k}
\end{equation}
The instability boundaries are shown in Fig. \ref{m_sigc}.

\section{Conclusions}
We have examined the dependence of the instability threshold of the harmonic oscillator with multiplicative noise on the mass of the Brownian particle.  Even in the white noise limit, the dependence is very different for the cases of random damping as opposed to random stiffness.  In the former case, the threshold increases as the square-root of the mass whereas in the latter case it is independent.  The situation is even more intricate when dichotomous noise in consider.  Then, the instability threshold is monotonically increasing for the case of random damping.  For random stiffness, the dependence is non-monotonic for small damping, first rising for small mass and then decreasing, while it is monotonically decreasing for large damping. For the case of random dichotomous variation of the mass, the dependence is always monotonically decreasing.  The stability criteria for the various cases were presented explicitly and their asymptotic limits discussed.

 The
distribution function for the energy was computed numerically and was seen to exhibit a  power-law cut-off at a value that increases exponentially in time.  This is connected to the 
fact that the system has no nonlinear saturation, and so time serves as the only cutoff.  It is also related to the fact that the instabilities of the asymptotic power-law
distributions of $<E>,$ $<E^{2}>$ and $<E^{3}>$, etc. occur at smaller and smaller noise thresholds.  An analytic calculation of the long-time asymptotic behavior of the distribution is an interesting open problem.


\begin{thebibliography}{99}
\bibitem{west} B. West and V. Seshadri, J. Geophys. Res. \textbf{86},
4293 (1981).

\bibitem{git1} M. Gitterman, Phys. Rev. E \textbf{70}, 036116 (2004).

\bibitem{25} A. Onuki, J. Phys.: Condens. Matter \textbf{9}, 6119 (1997).

\bibitem{26} J. M. Chomaz and A. Couairon, Phys. Fluids \textbf{11},
2977 (1999).

\bibitem{28} F. Hestol and A. Libchaber, Phys. Scr. \textbf{T9}, 126 (1985).

\bibitem{29} A. Saul and K. Showalter in \textit{Oscillations and Traveling
Waves in Chemical Systems}, R. J. Field and M. Burger, eds., (Wiley, New York, 1985).

\bibitem{30} M. Gitterman, B. Ya. Shapiro and I. Shapiro,  Phys. Rev.
B \textbf{65}, 174510 (2002).

\bibitem{ishi} A. Ishimaru, \textit{Wave Propagation and Scattering in Random
Media}, (IEEE Press, Piscataway, NJ, 1997).

\bibitem{kubo} R. Kubo in \textit{Stochastic Processes in Chemical Physics},  K. E. Shuler, ed. (Wiley, New York, 1969).

\bibitem{phil} O. M. Phillips, \textit{The Dynamics of the Upper Ocean }%
(Cambridge University Press,  Cambridge, 1977).

\bibitem{tur} M. Turelli,  \textit{Theoretical Population Biology }[Academic,
New York, 1977].

\bibitem{tak} H. Takayasu, A.--H. Sato, and M. Takayasu, Phys. Rev.
Lett. \textbf{79}, 966 (1997).


\bibitem{gi} M. Gitterman, 2010, J. Phys. C \textbf{248}, 012049 (2010).
\bibitem{gi1} M. Gitterman and I. Shapiro,  J. Stat. Phys., \textbf{144}, 139 (2011).
\bibitem{gi2}M. Gitterman, 4th Chaos Conference Dubna, Proceedings, unpublished (2011).
\bibitem{gi3}M. Gitterman, J. Modern Phys. \textbf{2}, 1136 (2011).




\bibitem{lam} R. Lambiotte and M.\ Ausloos,  Phys. Rev. E\textbf{\ 73},
011105 (2006).

\bibitem{gad} A. Gadomski A. and J. Si\'{o}dmiak,  Cryst. Res. Technol. 
\textbf{37}, 281 (2002).

\bibitem{rub} J. M. Rub\`{\i}  and A. Gadomski,  Physica A\textbf{\ 326},
333 (2003).

\bibitem{gad1} A. Gadomski, J. Si\'{o}dmiak, I. Santamar\`{\i}a-Holek, J.
M. Rub\`{\i},
and M. Ausloos, Acta Phys. Pol. B \textbf{36}, 1537 (2005).

\bibitem{per} A. T. P\'{e}rez , D. Saville, and C. Soria,  Europhys.
Lett.\textbf{\ 55}, 425 (2001).

\bibitem{gol} I. Goldhirsch  and G. Zanetti,  Phys. Rev. Lett. \textbf{70%
}, 1619 (1993).

\bibitem{lud} S. Luding  and H. J. Herrmann,  Chaos \textbf{9}, 673 (1999).

\bibitem{see} I. Temizer, M.Sc. thesis, University of California, Berkeley,
unpublished (2003), available at www.me.berkeley.edu/compmat/ilkerDOCS/MSthesis.pdf.

\bibitem{benz} W. Benz,  Spatium \textbf{6}, 3 (2000).

\bibitem{blum} J. Blum, et al.,Phys. Rev. Lett. \textbf{85}, 2426 (2000);
J. Blum  and G. Wurm,  Icarus \textbf{143}, 138 (2000).

\bibitem{weid} S. J. Weidenschilling, D. Spaute, D. R. Davis, F. Marzari,
and K. Ohtsuki,  Icarus \textbf{128}, 429 (1997).

\bibitem{kai} N. Kaiser,  Appl. Opt. \textbf{41}, 3053
 (2002).
 
\bibitem{nag} T. Nagatani,  J. Phys. Soc. Jpn. \textbf{65}, 3386 (1996).

\bibitem{benn} E. Ben-Naim , P. L. Krapivsky, and S. Redner, Phys. Rev.
E\textbf{\ 50}, 822 (1994).

\bibitem{aus} M. Ausloos  and K. Ivanova,  Eur. Phys. J. B \textbf{27},
177 (2002).

\bibitem{aus1} M. Ausloos  and K. Ivanova, in Proceedings of the Second
Nikkei Econophysics Symposium,  H. Takayasu,, ed.  (Springer Verlag, Berlin,
2004]).

\bibitem{yamada} T. Yamada and H. Fujisaka, Prog. Theor. Phys. \textbf{76}, 582 (1986).

\bibitem{Arnold} L. Arnold, \textit{Random Dynamics Systems}, (Springer Verlag, Berlin, 2003).






\bibitem{Kitahara} R. Kitahara, W. Horsthemke and R. Lefever, Phys. Lett. A \textbf{70}, 374 (1979).

\bibitem{KesBar} D. A. Kessler and E. Barkai, Phys. Rev. Lett. \textbf{105}, 120602 (2010).

\bibitem{Kubo}R. Kubo. J. Math. Phys. \textbf{4}, 174 (1963).
\end{thebibliography}
\end{document}